\documentclass[twocolumn,showpacs,preprintnumbers,amsmath,amssymb]{revtex4-1}

\usepackage{bm}
\usepackage{graphicx}

\begin{document}

%\preprint{}

%\title{Beyond the odd number limitation: Control matrix design for the time-delayed feedback control algorithm}
%\title{Design of time-delayed feedback control scheme  for stabilization of periodic orbits with an odd number of real unstable Floquet multipliers}
\title{Time-Delayed Feedback Control Design Beyond the Odd Number Limitation}


\author{Kestutis Pyragas and Viktor Novi\v{c}enko}
%\thanks{}
\affiliation{Center for Physical Sciences and Technology, A. Go\v{s}tauto 11, LT-01108 Vilnius, Lithuania}


\date{\today}

\begin{abstract}
We present an algorithm for a time-delayed feedback control design to stabilize periodic orbits with an odd number of positive Floquet exponents in autonomous systems. Due to the so-called odd number theorem  such orbits have been considered as uncontrollable by time-delayed feedback methods. However, this theorem has been refuted  by a counterexample and recently a corrected version of the theorem has been proved. In our algorithm, the control matrix is designed using a relationship between Floquet multipliers of the systems controlled by time-delayed and proportional feedback. The efficacy of the algorithm is demonstrated with the Lorenz and Chua systems.
\end{abstract}


\maketitle

Control of complex and chaotic dynamics is one of the central issues in applied nonlinear science. Starting with the work of Ott, Grebogi, and Yorke \cite{ogy90}, a variety of methods have been developed in order to stabilize unstable periodic orbits (UPOs) embedded in a chaotic attractor by employing tiny control forces \cite{sholl08}. A particularly simple and efficient scheme is time-delayed feedback control (TDFC) first introduced by one of us (KP) \cite{pyr92} and later extended or modified by different authors, e.g., \cite{soc94,*ahl04}. The TDFC has been successfully applied to many real-world problems in physical, chemical and biological systems ( c.f., e.g.,    \cite{sie08,*yam09}
and \cite{pyr06} for a review).

However, Nakajima \cite{nak97} has pointed out that time-delayed feedback schemes suffer from the so-called odd number limitation. The Nakajima's theorem states that unstable periodic orbits with an odd number of real Floquet multipliers (FMs) larger than unity cannot be stabilized by time-delayed feedback control. The limitation seemed to be supported by experimental and numerical evidence, and over the following years the research was focused on a search for various modifications of the TDFC in order to bypass the limitation \cite{schu97,*nak98,*pyr01,*pyr04,*pyr06a,*tam07,*hohn07}. Significant new knowledge has been gained ten years after the publication of the Nakajima's theorem, when Fiedler et al. \cite{fied07} have shown that the limitation is incorrect for autonomous systems.  The authors of Ref. \cite{fied07} considered a simple two-dimensional model system, a normal form for a subcritical Hopf bifurcation, which has a UPO with exactly one positive unstable Floquet multiplier, and showed  that it can be stabilized by the conventional TDFC scheme (see also Ref. \cite{just07}). The mechanism of stabilization identified by Fiedler et al. has been shown to work close to a subcritical Hopf bifurcation in a Lorenz system \cite{post07} and in a laser experiment \cite{schi11}. Similar results have been obtained for rotating waves near a fold bifurcation \cite{fied08}.

In all examples above, the choice of the structure of the control matrix is strongly related with the fact that the system is close to a bifurcation point. Though the odd number theorem is formally refuted for autonomous systems, there are no recipes for designing the control matrix far from bifurcation points. The aim of this letter is to fill this gap. Our research is mainly encouraged with the recent publication by Hooton and Amann \cite{hoo12} who presented a corrected version of the Nakajima's theorem for autonomous systems. We also use our recent results  based on a phase reduction theory extended for systems with time delay \cite{physd12,nov12} as well as a relationship between the Floquet multipliers of the systems controlled by  time-delayed and  proportional feedback \cite{pyr02}.

Let us consider an uncontrolled dynamical system $\dot{\mathbf{x}}(t) = \mathbf{f}\left(\mathbf{x}(t) \right)$ with $\mathbf{x}(t)\in \mathbb{R}^n$ and $\mathbf{f}: \mathbb{R}^n \rightarrow \mathbb{R}^n$ and assume that it has an unstable $T$-periodic orbit $\mathbf{x}(t)=\bm{\xi}(t)=\bm{\xi}(t+T)$, which we seek to stabilize by the time-delayed feedback control of the form
\begin{equation}
\dot{\mathbf{x}}(t) = \mathbf{f}\left(\mathbf{x}(t) \right)+\mathbf{K}[\mathbf{x}(t-\tau)-\mathbf{x}(t)],
\label{dfc}
\end{equation}
where $\mathbf{K}$ is an $n\times n$ control matrix and $\tau$ is a positive delay time. Provided that the delay time coincides with the period of the orbit, $\tau=T$, the periodic solution $\bm{\xi}(t)$ of the free system is also a solution of (\ref{dfc}) for any choice of the control matrix $\mathbf{K}$, i.e., the form (\ref{dfc}) yields a noninvasive control scheme. A necessary condition for the stability of the solution $\bm{\xi}(t)$ of the controlled system (\ref{dfc}) is given by the Hooton's and Amann's theorem \cite{hoo12}. To formulate this theorem let us assume that $\tau$ slightly differs from $T$. Then the controlled system (\ref{dfc}) has a periodic solution close to $\bm{\xi}(t)$ with a new period $\Theta$. Generally, the period $\Theta$ differs from $\tau$ and $T$; it is a function of $\mathbf{K}$ and $\tau$, $\Theta=\Theta( \mathbf{K},\tau )$, which satisfies $\Theta\left( \mathbf{K},T \right)=T$. The Hooton's and Amann's theorem claims, that the periodic solution $\bm{\xi}(t)$ is an unstable solution of the controlled system (\ref{dfc}) if the condition
\begin{equation}
(-1)^m \lim_{\tau \rightarrow T} \frac{\tau-T}{\tau-\Theta( \mathbf{K},\tau )}< 0
\label{teor}
\end{equation}
holds. Here $m$ is a number of real Floquet multipliers larger than unity for the periodic solution $\bm{\xi}(t)$ of the uncontrolled system. The criterion (\ref{teor}) differs from the Nakajima's version by the factor $\beta=\lim_{\tau\rightarrow T}(\tau-T)/(\tau-\Theta)$. It follows that the necessary (but not the sufficient) condonation for the TDFC to stabilize a UPO with an odd number $m$ is $\beta <0$. This condition predicts correctly the location of the transcritical bifurcation, which provides successful  stabilization of the UPO in the example of Fiedler et al. \cite{fied07,hoo12}.

The criterion (\ref{teor}) can be rewritten in a more handy form.
An explicit dependence of the factor $\beta$ on the control matrix $\mathbf{K}$ can be derived from (\ref{teor}) by expanding $\Theta$ in terms of a small mismatch $\tau-T$ up to the second order. This problem has been solved in our recent paper \cite{nov12} in a rather general formulation of a multiple-input multiple-output
system (c.f. \cite{just98} for the case of the scalar input). The approach used in \cite{nov12} is based on a phase reduction theory extended for systems with time delay \cite{physd12}. For the control law defined by (\ref{dfc}), the result of \cite{nov12} reads
\begin{equation}
\Theta(\mathbf{K},\tau)=T-(\tau-T)[\alpha(\mathbf{K})-1]+O[(\tau-T)^2],
\label{Theta}
\end{equation}
where $\alpha(\mathbf{K})$ is a coefficient that relates the phase response curve (PRC) $\mathbf{z}(t)$ of the periodic orbit of the controlled system (for $\tau=T$) with the PRC $\bm{\rho}(t)$ of the same orbit of the uncontrolled system, $\mathbf{z}(t)=\alpha(\mathbf{K})\bm{\rho}(t)$. The latter expression shows that the profile of the PRC of the controlled orbit is independent of the control matrix $\mathbf{K}$, only its amplitude $\alpha(\mathbf{K})$ depends on $\mathbf{K}$.  The PRC of the uncontrolled system can be computed as a $T$-periodic solution of the adjoint equation
\begin{equation}
\dot{\bm{\rho}}^T(t)=-\bm{\rho}^T(t)D\mathbf{f}\left(\bm{\xi}(t) \right)
\label{adj}
\end{equation}
for which the condition $\bm{\rho}^T(t) \dot{\bm{\xi}}(t)=1$  holds for any $t$.  Here the superscript ``T'' denotes the transpose operation and $D\mathbf{f}\left(\bm{\xi}(t) \right)$ is the Jacobian matrix of the uncontrolled system estimated on the periodic orbit.  Substituting (\ref{Theta}) into (\ref{teor}) we obtain a simple relationship between the factor $\beta$ and coefficient $\alpha$: $\beta=\alpha^{-1}$. The coefficient $\alpha$ has been estimated in \cite{nov12} so that for the factor $\beta$ we get
\begin{equation}
\beta=\alpha^{-1}(\mathbf{K})=1+\sum_{ij}^{n}\nolimits K_{ij}C_{ij},
\label{alpha}
\end{equation}
where $K_{ij}$ is the $i,j$ element of the control matrix and $C_{ij}=\int_0^T \rho_i(t) \dot{\xi}_j(t)dt$.
Here $\dot{\xi}_j(t)$ denotes the $j$-th component of derivative of the periodic orbit and $\rho_i(t)$ is the $i$-th component of the PRC of the uncontrolled orbit. Relation (\ref{alpha}) expresses explicitly the dependence of the factor $\beta$ on  the control matrix. To compute the coefficients $C_{ij}$ we need to solve Eq. (\ref{adj}). An algorithm for solution of this equation is described in Ref. \cite{physd12}; it requires a knowledge of at least one control matrix that provides the successful stabilization of the target UPO. Below we describe another way of estimating the  coefficients $C_{ij}$, without recourse to the solution of Eq. (\ref{adj}).

Note that the phase reduction theory identifies perfectly the transcritical bifurcation in the Fiedler et al. example \cite{fied07}. When the delay-induced orbit coalesces with the target UPO the trivial Floquet multiplier $\mu=1$ becomes degenerate. At the bifurcation point ($\beta=\alpha^{-1}=0$) the amplitude $\alpha$ of the PRC $\mathbf{z}(t)=\alpha \mathbf{\rho}(t)$ of the controlled orbit tends to infinity, i.e., the phase of the system becomes extremely sensitive to external perturbations.

In what follows, we present a practical recipe for designing the control matrix when a target UPO of dynamical system has a single $m=1$ real FM larger than unity. Any control matrix can be written in the form $\mathbf{K}=\kappa\mathbf{B}$, where $\kappa$ is a scalar control gain and $\mathbf{B}$  is a matrix with at least one element equal to $-1$ or $1$ and other elements in the interval $[-1,1]$. We can satisfy the Hooton's and Amann's necessary condition $\beta<0$ for any given matrix $\mathbf{B}$ if choose the control gain as
\begin{equation}
\kappa > \kappa^* \equiv -\left(\sum_{ij}^n\nolimits B_{ij} C_{ij}\right)^{-1}.
\label{thresh}
\end{equation}
However, this condition is not sufficient for the successful control. Without loss of the generality we assume that the threshold $\kappa^*$ is positive, since this can be always achieved  by appropriate choice of the sign of the matrix $\mathbf{B}$. We obtain  additional conditions for $\mathbf{B}$  by using a relationship between the Floquet multipliers of the TDFC and proportional feedback control (PFC) systems \cite{pyr02}. Consider the PFC problem derived from Eq. (\ref{dfc}) by replacing the time-delay term $\mathbf{x}(t-\tau)$ with $\bm{\xi}(t)$ and representing the control matrix as $\mathbf{K}=g\mathbf{B}$
\begin{equation}
\dot{\mathbf{x}}(t) = \mathbf{f}\left(\mathbf{x}(t) \right)+g\mathbf{B}[\bm{\xi}(t)-\mathbf{x}(t)].
\label{prop}
\end{equation}
The scalar $g$ defines the feedback gain for the PFC system. The problem of stability of the periodic orbit controlled by proportional feedback is relatively simple.
Small deviations $\delta \mathbf{x}(t)=\mathbf{x}(t)-\bm{\xi}(t)$ from the periodic orbit can be decomposed into eigenfunctions according to the Floquet theory $\delta \mathbf{x}(t)=\exp(\Lambda t)\mathbf{u}(t)$, where $\Lambda$ is the Floquet exponent (FE), and the $T$-periodic Floquet eigenfunction $\mathbf{u}(t)$ satisfies
\begin{equation}
\dot{\mathbf{u}}(t)+\Lambda \mathbf{u}(t) = \left[ D\mathbf{f}\left(\bm{\xi}(t) \right)-g \mathbf{B}\right] \mathbf{u} (t).
\label{fe_b}
\end{equation}
This equation produces $n$ FEs $\Lambda_j$, $j=1,\ldots,n$ [or FMs $\exp(\Lambda_j T)$]. The Floquet problem for the TDFC system (\ref{dfc}) is considerably more difficult, since it is characterized by an infinity number of FEs. Let us denote the FEs of the periodic orbit controlled by time-delayed feedback by $\lambda$ and the corresponding FMs by $\mu=\exp(\lambda T)$. The Floquet eigenvalue problem for the TDFC system can be presented in a form of Eq. (\ref{fe_b}) with the following replacement of the parameters: $\Lambda \rightarrow \lambda$ and $g \rightarrow \kappa [\exp(-\lambda T)-1]$. Provided the FM $\exp(\lambda T)$ is real valued, this property leads to  the following parametric equations (c.f. \cite{pyr02})
\begin{equation}
\lambda = \Lambda(g), \quad
\kappa = g\left[1-\exp\left(-\Lambda(g) T\right)\right]^{-1},
\label{par}
\end{equation}
which allow a simple reconstruction of the dependence $\lambda=\lambda(\kappa)$ for some of branches of  FEs of the TDFC system using the knowledge of the similar dependence $\Lambda=\Lambda(g)$ for the PFC system. Though Eqs. (\ref{par}) are valid only for the real valued FMs, it appears that exactly these branches are most relevant for the stability of the TDFC system.

\begin{figure}[t!]
\centering\includegraphics[width=0.45\textwidth]{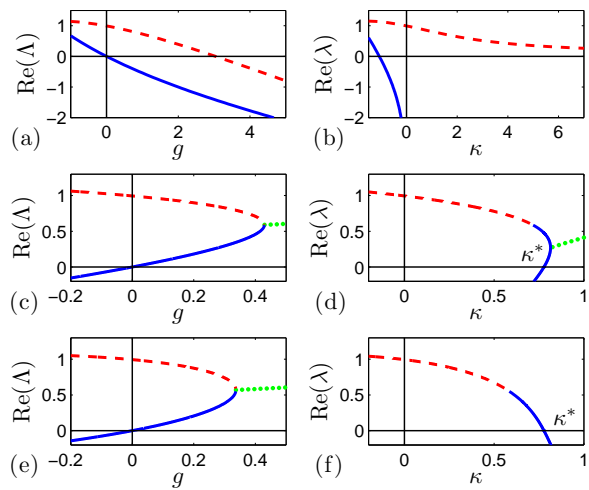}
\caption{\label{fig1} (Color online) Three typical scenarios for the dependence of the FEs of the Lorenz system on the feedback gain for PFC (left-hand column) and TDFC (right-hand column) at differen matrixes $\mathbf{B}$: (a) and (b) $[0,0,0;0,1,0;0,0,0]$; (c) and (d) $[0,0,0;-1,0,0.3;0,0,0]$; (e) and (f) [0,0,0;-1,0,0.5;0,0,0]. In (a), (c) and (e), blue solid and red dashed curves represent trivial and unstable FEs (both real valued) for PFC, respectively.   In (b), (d) and (f), the corresponding curves show reconstructed values of FEs for TDFC.  $\kappa^*$ is a threshold  control gain, where the trivial FE branch crosses zero in the PFC system. Green dotted curves in (c) and (e) show real parts of complex conjugate FEs, which cannot be transformed to TDFC by Eqs. (\ref{par}). Green dotted curve in (d) shows the real part of complex conjugate FEs emerged from coalescence of two real FEs of the TDFC system; it is computed by DDE-BIFTOOL \cite{biftool}.}
\end{figure}

To demonstrate the advantages of Eqs. (\ref{par}) we refer to the Lorenz system described by the state vector $\mathbf{x}(t)=[x_1(t),x_2(t),x_3(t)]^T$ and the vector field
\begin{equation}
\mathbf{f}(\mathbf{x})=[10(x_2-x_1),x_1(28 - x_3)-x_2,x_1x_2-8/3x_3]^T.
\label{lor}
\end{equation}
We take the standard values of the parameters, which produce the classical chaotic Lorenz attractor and consider the stabilization of its symmetric period-one UPO  with the period $T \approx 1.559$ and the single unstable FM $\mu \approx 4.713$. In Fig. \ref{fig1} we show three typical dependencies of the FEs on the coupling strength for the PFC $\Lambda=\Lambda(g)$ (left-hand column) and the TDFC $\lambda=\lambda(\kappa)$ (right-hand column) systems obtained with different matrixes $\mathbf{B}$. The dependencies $\Lambda=\Lambda(g)$ for the PFC are derived from Eq. (\ref{fe_b}). We plot only two branches of the FEs, originated from the unstable FE of the free system (red dashed curve) and from the trivial FE (blue solid curve crossing the origin). The branch corresponding to the negative FE of the free system does not influence the stability of the TDFC. The dependencies $\lambda=\lambda(\kappa)$ for the TDFC are obtained using the transformation (\ref{par}). We see that the case (a)-(b) provides successful control for the PFC but it is unsuccessful for the TDFC. The case (c)-(d) is again unsuccessful for the TDFC; here two real FEs coalesce in the positive region and produce a pair of complex conjugate FEs with the positive real part, which grows with the increase of $\kappa$. Finally, the case (e)-(f) is potentially successful for the TDFC; here the branch of unstable FE  (which results from two branches of the PFC system) decreases monotonically with the increase of $\kappa$ and becomes negative for $\kappa>\kappa^*$.

Now we show that the threshold $\kappa^*$ obtained from the FEs of the PFC system and transformation (\ref{par}) coincides with the definition (\ref{thresh}) derived from the Hooton's and Amann's criterion. The values  $\lambda(\kappa)$ of the TDFC system with $\kappa$ close to the threshold $\kappa^*$ result from the values of the trivial FE $\Lambda(g)$ of the PFC system with $g$ close to zero. To derive an expression for $\kappa^*$ we expand the dependence $\Lambda(g)$ for the trivial FE in Taylor series
\begin{equation}
\Lambda(g)T = ag+bg^2+O(g^3).
\label{tayl}
\end{equation}
Substituting (\ref{tayl}) into (\ref{par}) and taking the limit $g\rightarrow0$ we get $\kappa^*=a^{-1}$. An expression for the coefficient $a$ can be derived by applying the perturbation theory to Eq. (\ref{fe_b}). To this end we write the trivial eigenmode
in the form $\mathbf{u}(t)=\mathbf{u}_0(t)+g\mathbf{u}_1(t)
+O(g^2)$. Substituting this expansion and (\ref{tayl}) into (\ref{fe_b}),  we get in zero approximation  $\dot{\mathbf{u}}_0(t)=D\mathbf{f}\left(\bm{\xi}(t) \right) \mathbf{u}_0(t)$. The solution of this equation is  $\mathbf{u}_0(t)=\dot{\bm{\xi}}(t)$. In the first order approximation, we obtain
\begin{equation}
\dot{\mathbf{u}}_1(t)=D\mathbf{f}\left(\bm{\xi}(t) \right)\mathbf{u}_1(t)-\left( \mathbf{B} +\mathbf{I} a/T\right)\dot{\bm{\xi}}(t),
\label{first}
\end{equation}
where $\mathbf{I}$ is the identity matrix. Multiplying Eq. (\ref{first}) on the LHS by $\bm{\rho}^T(t)$ and summing it with Eq. (\ref{adj}) multiplied on the RHS by $\mathbf{u}_1(t)$, we get:
\begin{equation}
\frac{d}{dt}\left(\bm{\rho}^T(t) \mathbf{u}_1(t)  \right)=-\bm{\rho}^T(t)\left( \mathbf{B} +\mathbf{I} a/T\right)\dot{\bm{\xi}}(t).
\label{fora}
\end{equation}
Finally, we integrate this equation over the period $T$ and  obtain  $a=-\int_0^T\bm{\rho}^T(t)\mathbf{B}\dot{\bm{\xi}}(t)dt$, which means that the value $a^{-1}$ coincides with the threshold $\kappa^*$ defined in (\ref{thresh}).

A relation of the coefficient $a$ with the matrix $\mathbf{B}$
\begin{equation}
a = -\sum_{ij}^n\nolimits B_{ij} C_{ij}
\label{a}
\end{equation}
provides an alternative way to estimate the coefficients $C_{ij}$. The particular coefficient $C_{\tilde{i}\tilde{j}}$ can be estimated as $C_{\tilde{i}\tilde{j}}=-a$ if we choose the matrix $\mathbf{B}$ with all zero elements except for $B_{\tilde{j}\tilde{i}}=1$. Then the coefficient $a$ in expansion (\ref{tayl}) can be obtained by numerical computation of the dependence $\Lambda(g)$  for small $g$.

Apart from the the Hooton's and Amann's condition (\ref{thresh}), the successful control requires that the derivative $d\lambda/d\kappa$ at the threshold $\kappa=\kappa^*$ to be negative  [see Fig. \ref{fig1}(f)]. Substituting (\ref{tayl}) into (\ref{par}) we get
\begin{equation}
\left. \frac{d\lambda}{d\kappa} \right|_{\kappa=\kappa^*} =\lim_{g\rightarrow 0} \frac{d\Lambda /dg}{d\kappa /dg}=\frac{2a}{T(1-2b/a^2)}<0.
\label{ine}
\end{equation}
The parameter $a$ is positive by assumption of the positiveness of $\kappa^*$. Then this condition simplifiers  to
\begin{equation}
1-2b/a^2<0.
\label{ine2}
\end{equation}
By extending the above perturbation theory for Eq. (\ref{fe_b}) up to the second order terms with respect to $\kappa$, we derive the following expression for the coefficient $b$:
\begin{equation}
\label{forb}
b=-\frac{a}{T}\int_0^T\bm{\rho}^T(t)\mathbf{u}_1(t)dt- \int_0^T\bm{\rho}^T(t)\mathbf{B}\mathbf{u}_1(t)dt.
\end{equation}
This allows us to write the relation of the coefficient $b$ with the matrix $\mathbf{B}$ in the quadratic form
\begin{equation}
\label{forb2}
b=\sum_{ijkl}^n\nolimits B_{ij}B_{kl}D_{ijkl}
\end{equation}
with coefficients $D_{ijkl}=D_{klij}$. These coefficients can be obtained in a similar way as  the coefficients $C_{ij}$ by taking specific forms of the matrix $\mathbf{B}$ and estimating $b$ from the dependence $\Lambda(g)$ of the trivial FE  for small $g$.

The knowledge of the coefficients $C_{ij}$ and $D_{ijkl}$ allows an explicit computation of the parameters $a$ and $b$ for any given matrix $\mathbf{B}$. As a result we can simply verify the condition (\ref{ine2}) and estimate the threshold $\kappa^*$ in (\ref{thresh}).

Finally, we can summarize our algorithm as follows: (i) choose the structure of the matrix $\mathbf{B}$ with only several nonzero elements in such a way as to make possible the coalescence of the positive and trivial Floquet branches of the PFC system [like in Fig. (\ref{fig1}) (c) or (e)]; (ii) for the given structure of the matrix $\mathbf{B}$, estimate the relevant coefficients $C_{ij}$ and $D_{ijkl}$; (iii) choose the values of nonzero elements of the matrix $\mathbf{B}$ such as to satisfy condition (\ref{ine2}); (iv) compute the threshold $\kappa^*$ and satisfy condition (\ref{thresh}). Note that our algorithm considers only most important branches of the FEs and its final outcome has to be verified  by more detailed analysis of the stability of the TDFC system. Nevertheless, the algorithm gives a simple practical recipe for the selection of appropriate control matrixes and  works well for typical chaotic systems.

\begin{figure}[t!]
\centering\includegraphics[width=0.44\textwidth]{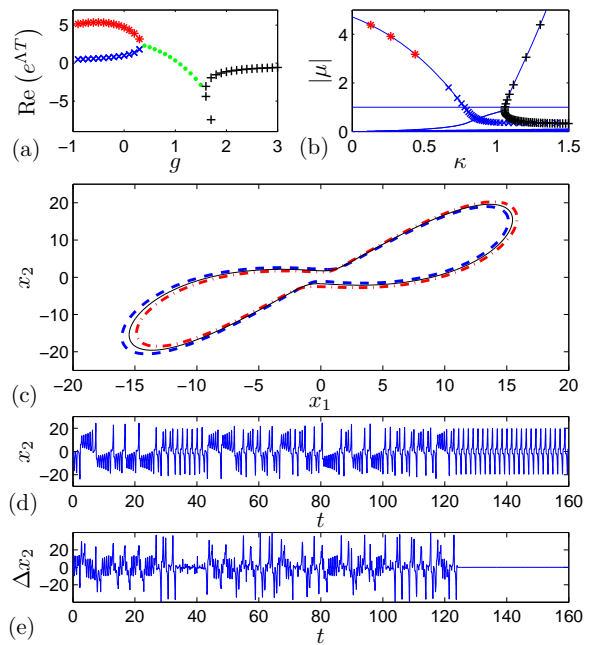}
\caption{\label{fig2} (Color online) Stabilization of the period-one UPO of the Lorenz system with the matrix $\mathbf{B}=[0, 0, 0;-1, 0, 0.5;0, 0, 0]$. (a) FMs vs. $g$ for the PFC.  Blue crosses and red asterisks represent the trivial and unstable branches, respectively (they are real-valued). Green points represent the real part of complex conjugate FMs. Black "plus signs" show a new pair of real-valued branches appeared from the complex conjugate FMs. (b) Absolute values of FMs vs. $\kappa$ for the TDFC. Solid curves are obtained by the DDE-BIFTOOL, while symbols show the reconstruction of the FMs from panel (a) via Eqs. (\ref{par}). Both branches (one marked by asterisks and crosses and another by "plus signs") that define the stability interval $\kappa\in[0.78, 1.06]$ of the TDFC are reconstructed from the PFC system. (c) $(x_1,x_2)$ projection of periodic orbits. Blue dashed and red dash dotted curves show the stable delay-induced orbit for $\kappa=0.63$ before the trascritical bifurcation ($\kappa^*\approx 0.78$) and the unstable delay-induced orbit for $\kappa=1.05$ after the bifurcation, respectively. The target orbit is presented by black solid curve.  (d) and (e) Dynamics of $x_2(t)$ and difference $\Delta x_2(t)=x_2(t)-x_2(t-\tau)$ for $\kappa=0.865$ and filter parameters $\tau_w=0.5$ and $\varepsilon=2$.
}
\end{figure}

First we discuss the details of application of our algorithm for  the Lorenz system (\ref{lor}). Motivated by a ``common sense'' assumption we started our analysis with the diagonal matrix $\mathbf{B}$. However, it appeared that such a choice, which works well for PFC systems, does not satisfy the first point of our algorithm. The impossibility to attain successful control with the diagonal control matrix can probably explain why the Lorenz system has not been stabilized by a conventional TDFC until now. We found that the  requirements of our algorithm can be satisfied by many different nondiagonal configurations of the  matrix $\mathbf{B}$. Here we show the results with the matrix $\mathbf{B}$ that has only two nonzero elements $B_{21}=-1$ and $-1\leq B_{23} \leq 1$. The relevant coefficients for such a matrix configuration are: $C_{21}\approx 1.286$, $C_{23} \approx 1.5\times 10^{-3}$, $D_{2121}\approx 0.163$, $D_{2323} \approx 3.792$ and $D_{2123}=D_{2321} \approx 9.7\times 10^{-8}$. The inequality (\ref{ine2}) leads to the requirement $|B_{23}|>0.418$. We choose $B_{23}=0.5$ and obtain the threshold $\kappa^* \approx 0.78$. As is seen from Fig. \ref{fig2}, these estimates predict correctly the successful control. In panels (a) and (b) we compare the values of FMs of the TDFC system reconstructed from the PFC system with those obtained via direct analysis of the TDFC system by the DDE-BIFTOOL package \cite{biftool}. Surprisingly, Eqs. (\ref{par}) allow us to obtain not only the threshold $\kappa^*$, but also the interval of stability of the controlled orbit, since the branch of FMs (marked by ``plus signs'') that defines the loss of the stability is reconstructed from the PFC system as well. The stabilization of the UPO  at the threshold $\kappa^*$ is caused by  transcritical bifurcation as well as in the example of Fiedler et al.  \cite{fied07}. The delay-induced periodic orbits in vicinity of the bifurcation point are shown in panel (c). Finally, panels (d) and (e) show the dynamics of the controlled system obtained by integration of Eqs. (\ref{dfc}) and (\ref{lor}) \cite{mis}. To reduce the transient time, the moment of switching on the control has been determined by a filter equation $\dot{w}=\left\lbrace|x_1(t)-x_1(t-\tau)|-w(t)\right\rbrace/\tau_w$ \cite{pyr09}. The filter estimates the closeness of the system state to the UPO and  the control is activated only when the variable $w$ becomes small, $w(t)<\varepsilon$.

\begin{figure}[t!]
\centering\includegraphics[width=0.45\textwidth]{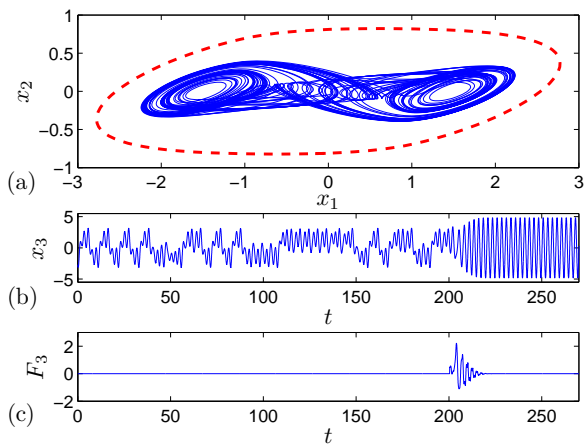}
\caption{\label{fig3} (Color online) Stabilization of a UPO in the Chua system with $\mathbf{B}=[0, 0, 0;0, 0, 0;1, 0, 0.3]$. (a) The $(x_1,x_2)$ projection of a chaotic attractor (blue solid curve) and the target UPO (red dashed curve). (b) and (c) Dynamics of  $x_3(t)$ and the third component of the TDFC force $F_3(t)$. The control is switched on at $t=200$ with  $\kappa=1.2$.}
\end{figure}

To demonstrate the universality of our approach  we refer to another example, the   Chua system \cite{chua07} defined by the state vector $\mathbf{x}=[x_1,x_2,x_3]^T$ and  the vector field
\begin{equation}
\label{chua}
\mathbf{f}(\mathbf{x})=[9\left( x_2-\phi(x_1) \right), x_1-x_2+x_3,-100/7 x_2]^T,
\end{equation}
where $\phi(x_1)=2/7x_1-3/14( |x_1+1|-|x_1-1|)$. The $(x_1,x_2)$ projection of a chaotic trajectory and the target UPO of the system are shown in Fig. \ref{fig3} (a). Here the target UPO is outside of the strange attractor; its period is $T\approx 2.483$ and the single unstable FM $\mu=2.325$. We choose a nondiagonal configuration of the matrix $\mathbf{B}$ with two nonzero elements $B_{31}=1$ and $-1\leq B_{33} \leq 1$. Then the relevant coefficients are  $C_{31}\approx-2.02$, $C_{33}\approx 3.01$, $D_{3131} \approx 2.46$, $D_{3333} \approx 1.85$, and $D_{3133}=D_{3331} \approx -2.21$. For  $B_{33}=0.3$, the inequality (\ref{ine2}) is satisfied and the threshold value of the control gain is $\kappa^* \approx 0.89$.  The successful stabilization of the UPO is demonstrated in panels (b) and (c)  for $\kappa=1.2$.

In conclusion, we have presented a practical recipe for time-delayed feedback control design, which enables the stabilization of periodic orbits with an odd number of real Floquet multipliers larger than unity. The algorithm is suited for autonomous systems far from bifurcation points of periodic orbits. Using this algorithm we managed to stabilize the periodic orbits in the Lorenz and Chua systems, which have been considered as classical examples unaccessible for the conventional time-delayed feedback control. Our findings will extend the possibilities for further implementations of time-delayed feedback control in practical applications.

This research was funded by the European Social Fund under the Global Grant measure (grant No.~VP1-3.1-\v{S}MM-07-K-01-025).

\bibliography{references}

\end{document}